# Structural transition-driven antiferromagnetic to spin-glass transition in Cd-Mg-Tb 1/1 approximants

Farid Labib[1], Daisuke Okuyama[1], Nobuhisa Fujita[1], Tsunetomo Yamada[2], Satoshi Ohhashi[1], Daisuke Morikawa[1], Kenji Tsuda[3], Taku J. Sato[1] and An-Pang Tsai[1]

[1] Institute of Multidisciplinary Research for Advanced Materials (IMRAM), Tohoku University, Sendai 980-8577, Japan
[2] Faculty of Science, Department of Applied Physics, Tokyo University of Science, Katsushika-ku, Tokyo, Japan
[3] Frontier Research Institute for Interdisciplinary Sciences, Tohoku University, Sendai 980-8578, Japan

E-mail: labib.farid.t2@dc.tohoku.ac.jp

**Abstract**

The magnetic susceptibility of the quasicrystal 1/1 approximants in a series of $Cd_{85-x}Mg_xTb_{15}$ ($x$ = 5, 10, 15, 20) alloys was investigated in detail. The occurrence of antiferromagnetic (AFM) to spin-glass (SG)-like transition was noticed by increasing Mg content of the compounds. Transmission electron microscopy analysis evidenced a correlation between the magnetic transition and suppression of the monoclinic superlattice ordering with respect to the orientation of the $Cd_4$ tetrahedron at $T$ > 100 K. The possible origins of this phenomenon were discussed in detail. The occurrence of the AFM to SG-like magnetic transition is associated with the combination of chemical disorder due to a randomized substitution of Cd with Mg and the orientational disorder of the $Cd_4$ tetrahedra.

## 1. Introduction

Quasicrystals (QC) have been an object of research since their initial discovery in 1984 [1]. They are characterized as solid state materials possessing a particular kind of long-range order without periodicity in their atomic configuration [2]. Tsai-type icosahedral quasicrystals (iQCs), in particular, exhibiting an icosahedral point symmetry consist of rare earth (RE) elements along with Cd or a mix of other two metallic species following the substitution rules [3–5]. Their atomic structure composed of identical clusters with five successive shells [6–8]; as seen in Figure 1(a). Generally, their temperature-dependence magnetization at high temperatures (except RE = Yb, which is divalent in the QCs) follows the Curie-Weiss law with effective magnetic moments being consistent with free $RE^{3+}$ ions. At low temperatures, spin-glass (SG)-like freezing behavior has been reported for QCs [9–12].

In the near proximity of iQCs in some alloy systems, in terms of chemical composition, there exist compounds that contain the same rhombic triacontahedron (RTH) clusters being arranged in a body-centred cubic (BCC) packing rather than in aperiodic manner [13,14]. These compounds are denoted as approximant crystals (APs). The configuration of RE atoms within the unit cell of the Tsai-type 1/1 AP is illustrated schematically in Figure 1(b). Amongst the known 1/1 APs, the binary $Cd_6RE$ [11,15–17], ternary Au-Al-RE and Au-SM-RE systems (RE = rare earth, SM = Si, Ge or Sn) [18–20] have received considerable attention since they exhibit a long-range magnetic ordering at low temperatures. The binary $Cd_6RE$ compounds, in particular, exhibit an antiferromagnetic (AFM) ordering evidenced by one or several anomalies in their magnetic susceptibilities [15,16].

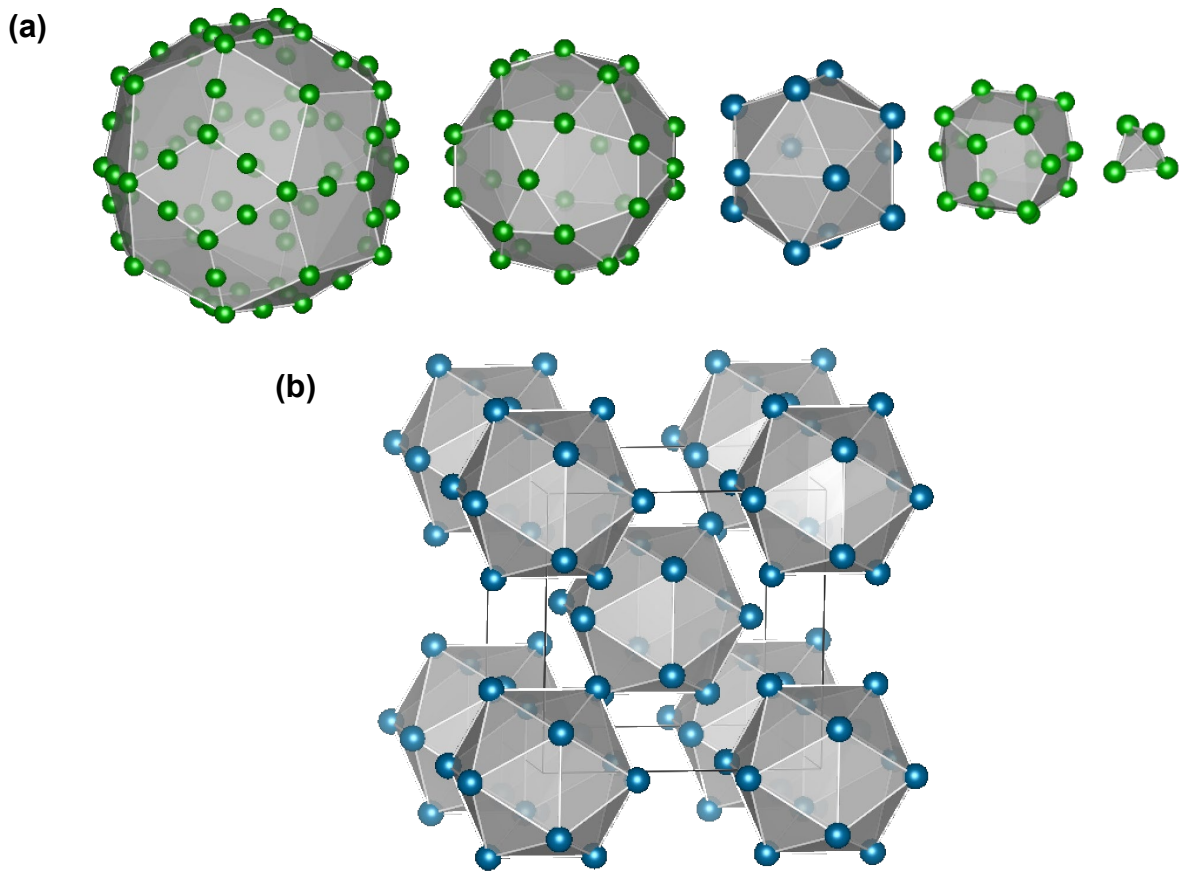

**Figure 1.** (a) Shell structure of the Tsai-type icosahedral quasicrystal (iQC). From left to right: Cd rhombic triacontahedron (RTH) (92 atoms), Cd icosidodecahedron (30 atoms), Rare Earth (RE) icosahedron (12 atoms), Cd dodecahedron (20 atoms), and inner Cd tetrahedron (4 atoms). (b) RE sites on the icosahedron shell of the RTH clusters in the atomic structures of the 1/1 approximant (AP) in $Cd_6Tb$ compound.

One of the unique features of the compounds in the Cd–RE systems is that they show high solubility of Mg atoms that locate on Cd sites due to their almost similar atomic size and valence. This leads to extension of the single-phase iQC and 1/1 AP in the ternary Cd–Mg–RE phase diagrams [21,22]. For example, the single phase 1/1 AP in Cd-Mg-Tb system spans up to $Cd_{65}Mg_{20}Tb_{15}$. This paper, therefore, aims to study the consequences of Mg addition on the magnetic properties of 1/1 APs in the Cd–Mg–Tb system especially focusing on the relevance of minor changes in the crystal structure. The AFM to SG-like transition is noticed to be triggered by increasing Mg content of the alloys. Such magnetic transition is associated with order-disorder-type phase transition with respect to the orientation of the central $Cd_4$ tetrahedron. The results indicate that Mg addition effectively suppresses the ordering of $Cd_4$ tetrahedra, which is a key element in the occurrence of AFM magnetic transition.

## 2. Experiment

Polycrystalline samples of $Cd_{85-x}Mg_xTb_{15}$ ($x$ = 5, 10, 15, 20) 1/1 APs were prepared by loading a total of three grams of pure Cd (99.99%), Mg (99.9999%) and Tb (99.9999%) into stainless-steel tubes. The tubes were further sealed in quartz tubes under an Ar gas atmosphere (~550 Torr). After an initial melting of the prepared alloys at 973 K employing arc-welding, they were isothermally annealed at 773 K for 100 h. In order to identify the phase quality of the resulting alloys, powder x-ray diffractometry (XRD) was performed using Cu-$K_\alpha$. In addition, the scanning electron microscopy (SEM; Hitachi SU-6600) equipped with energy dispersive x-ray (EDX) spectrometer was utilized to investigate the microstructure and local compositions of the prepared alloys. All the samples are single phase polycrystalline 1/1 APs, as described later. For electron backscatter diffraction (EBSD) analysis, the surface of the samples was polished by ion bombardment machine with accelerated voltage, gun current, milling time and ion angle of 2 kV, 2 mA, 3 h, and 12°, respectively. For transmission electron microscopy (TEM) analysis, the samples were crushed in ethanol and transferred to a Cu grid. An energy-filter TEM JEM-2010FEF was used at an accelerating voltage of 100 kV with a liquid-nitrogen cooling specimen holder.

The temperature dependence of the magnetic susceptibility was measured between 2 – 300 K under an external DC field of 100 Oe using superconducting quantum interfering device (SQUID) magnetometer (Quantum Design, MPMS-XL). Both zero-field cooled (ZFC) and field cooled (FC) data were collected. The AC magnetization of two representative samples were measured for $2 < T < 300$ K and frequencies varying from 1 to 100 Hz with $H_{AC} = 3$ Oe.

## 3. Results and discussion

Figure 2(a) shows a typical backscattered SEM image from a microstructure of the $Cd_{80}Mg_5Tb_{15}$ 1/1 AP annealed at 773 K for 100 h. The microstructure exhibits a homogeneous single phase. The EDX analysis of the specified region provided the composition values shown on the top right corner of the image, being consistent with the nominal values. Similar results were also obtained for the remaining samples, as listed in Table 1. The obtained data show no sign of Fe, O or C as a possible contamination from the synthesis procedure. Figure 2(b) represents electron backscatter diffraction (EBSD) Kikuchi pattern taken from the sample. As provided elsewhere [21,22], the deformed pentagonal shape of the Kikuchi bands surrounding the $[0\bar{3}2]$ pole, noticed by degenerate intersection of the Kikuchi bands as well as the occurrence of split diverging bands clearly differentiate the material from 2/1 AP and iQC, confirming that it is 1/1 AP.

Figure 3(a) provides powder XRD patterns of the $Cd_{85-x}Mg_xTb_{15}$ ($x$ = 5, 10, 15, 20) alloys after an isothermal annealing at 773 K for 100 h. The calculated pattern obtained from the refined structure of the binary $Cd_6Tb$ 1/1 AP with space group $Im\bar{3}$ [23] is co-plotted on the bottom of the same figure for comparison. Only high intensity peaks are indexed in the calculated pattern. The observed peak intensities for the four compounds are fairly consistent with the calculation. This indicates that the 1/1 AP phase is formed inside a relatively elongated compositional area up to ~ 20 at.% Mg. The variation of the lattice parameter as a function of analyzed Mg content of the prepared alloys is displayed in Figure 3(b). Clearly, the lattice parameter increases from 15.490(3) Å in $Cd_6Tb$ [23] to 15.527(3) Å in $Cd_{64.09}Mg_{21.34}Tb_{14.57}$. This is due to the slightly larger atomic radius of Mg (1.60 Å) compared to that of Cd (1.57 Å) [24] leading to expansion of the unit cell after the Mg substitutes Cd sites, as expected from Vegard's law. However, the lattice parameters do not fall on a single line but on two linear segments with different slopes meeting at around 13 at.% Mg, above which the unit cell undergoes a swift expansion. The crossover point might be correlated with the filling up of some preferred atomic sites for Mg, which will be discussed in detail later.

Figure 4(a) depicts the temperature dependence of DC magnetic susceptibility for the $Cd_{65}Mg_{20}Tb_{15}$ 1/1 AP, as a typical example, measured under 100 Oe in a temperature range of 2–300 K. Measured susceptibility data under the FC and ZFC conditions are shown with open and filled circles, respectively. The corresponding inverse magnetic susceptibility data shown in the inset fit well with the Curie-Weiss law in a high temperature region (100 K $< T <$ 300 K):

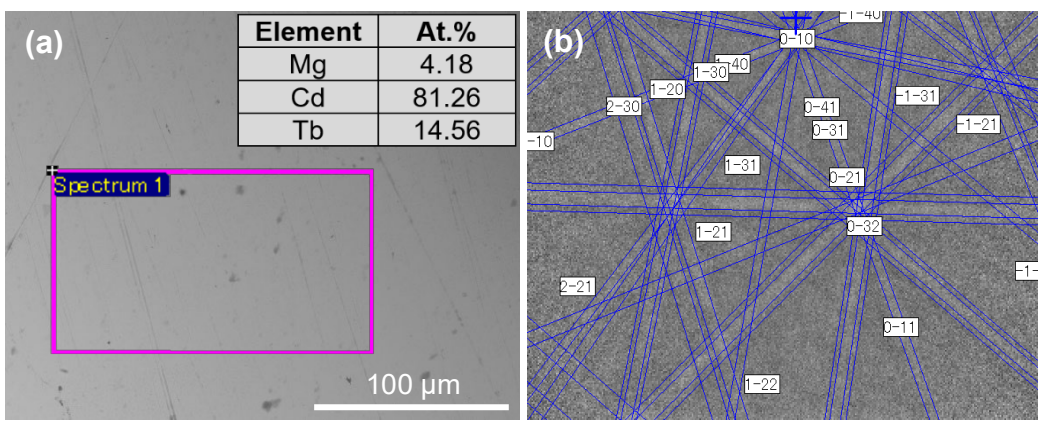

**Figure 2.** (a) Backscattered scanning electron microscopy (SEM) image. The energy dispersive x-ray (EDX) analysis of specific region is shown on the top right corner. (b) Electron backscatter diffraction (EBSD) Kikuchi pattern of the $Cd_{80}Mg_5Tb_{15}$ 1/1 AP. The pattern exhibits split Kikuchi bands forming distorted pentagonal bands with pseudo-five-fold [0-32] pole in the center suggesting a significant deviation of the 1/1 AP structure from an ideal iQC one.

**Table 1.** Nominal alloy compositions and analyzed compositions of the alloys examined in this study.

| Nominal composition | Analyzed compositions |
|---|---|
| $Cd_{80}Mg_5Tb_{15}$ | $Cd_{81.26}Mg_{4.18}Tb_{14.56}$ |
| $Cd_{75}Mg_{10}Tb_{15}$ | $Cd_{74.18}Mg_{11.53}Tb_{14.29}$ |
| $Cd_{70}Mg_{15}Tb_{15}$ | $Cd_{68.7}Mg_{16.51}Tb_{14.79}$ |
| $Cd_{65}Mg_{20}Tb_{15}$ | $Cd_{64.09}Mg_{21.34}Tb_{14.57}$ |

$$\chi(T) = \frac{N_A \mu_{eff}^2 \mu_B^2}{3k_B(T-\theta_w)} + \chi_0 \quad (1)$$

where $N_A$, $k_B$, $\mu_B$ and $\chi_0$ stand for Avogadro's number, the Boltzmann factor, Bohr magneton and the temperature-independent magnetic susceptibility, respectively. The $\chi_0$ is approximately zero in the present samples. The Weiss temperature, $\theta_w$, and the effective magnetic moment, $\mu_{eff}$, are estimated for each of the four compounds by linear least-squares fitting of their inverse susceptibility data at 100 – 300 K to the Curie-Weiss law and extrapolating the fitting data to the temperature axis. Table 2 lists the estimated $\mu_{eff}$, $\theta_w$, and $T_f$ or $T_N$ for samples possessing 5 and 20 at.% Mg. As given, the measured $\theta_w$ values are negative demonstrating AFM interactions between the magnetic moments. An uncertainty in the estimation of $\theta_w$ corresponds to the fitting over different temperature ranges. The $\mu_{eff}$ values are consistent with the ideal values calculated for the $Tb^{3+}$ free ion, i.e. 9.72 $\mu_B$, confirming that Tb ions are trivalent.

A magnetic phase transition temperature, at which an anomaly associated either with magnetic freezing ($T_f$) or AFM ordering ($T_N$) occurred, was determined from the magnified low-temperature DC susceptibility curves shown in Figure 4(b). The figure clearly evidences the occurrence of two anomalies in each of the susceptibility curves for the 1/1 APs containing 5, 10, 15 and 20 at.% Mg in Cd-Mg-Tb system. In $Cd_{80}Mg_5Tb_{15}$, for instance, both $\chi_{ZFC}$ and $\chi_{FC}$ curves are peaked at $T \sim 21$ K, while at $T \sim 5$ K only ZFC shows a maximum. The onset of the bifurcation between ZFC and FC data occurs at $T \sim 12$ K. The susceptibility curve of $Cd_{65}Mg_{20}Tb_{15}$ exhibits similar features to that of $Cd_{80}Mg_5Tb_{15}$ except the maximum of ZFC data that appears at $T \sim 13$ K. Below $T \sim 13$ K, the bifurcation between ZFC and FC magnetization curves takes place. Such a bifurcation points to a SG-like freezing transition [25].

Figure 5 depicts in-phase ($\chi'_{AC}$) and out-of-phase ($\chi''_{AC}$) components of the AC magnetic susceptibilities for the $Cd_{80}Mg_5Tb_{15}$ and $Cd_{65}Mg_{20}Tb_{15}$ alloys under selected frequencies between 1 to 100 Hz. The insets in Figures 5(a) and 5(b) provide magnified views of the susceptibility curves around the cusps. The cusps in $\chi'_{AC}$ curves are noticed at exactly the same temperatures as those observed in the DC magnetization curves in Figure 4(b). As seen from the insets, the amplitude and position of the cusp in the $Cd_{80}Mg_5Tb_{15}$ alloy (Figure 5(a), at around 21 K) are independent of the frequency of the applied field, whereas they are systematically affected by the frequency variation in the $Cd_{65}Mg_{20}Tb_{15}$ alloy (Figure 5(b), at around 13−14 K). At $T \sim 5$ K, however, both alloys exhibit frequency-dependent variation of the $\chi'_{AC}$. These results confirm that the $Cd_{80}Mg_5Tb_{15}$ alloy undergoes an AFM ordering and SG-like freezing at $T \sim 21$ and 5 K, respectively. This is consistent with the earlier reports about the binary $Cd_6Tb$ where the partial participation of Tb moments in the AFM ordering at $T = 24$ K and the SG freezing of the rest at lower temperature is reported [11,15,16,26–28]. The out-of-phase components $\chi''_{AC}$ evidence the occurrence of two frequency dependent peaks for both alloys: The strong peaks at $T \sim 5$ K followed by weak peaks at $T \sim 12-13$ K (indicated by arrows in Figures 5(c) and 5(d)). The weak peak at $T \sim 12$K in Cd$_{80}$Mg$_5$Tb$_{15}$, which disappears with the frequency increment, might reflect a certain, yet unclarified slow spin dynamics in the AFM regime. The Cd$_{65}$Mg$_{20}$Tb$_{15}$ alloy, on the other hand, exhibits two successive SG-like transitions at $T \sim$ 5 and 13.5 K. This implies that the Mg substitution to the Cd$_6$Tb 1/1 AP structure (shown in Figure 1) disrupts the long-ranged AFM ordering and induces the SG-like transition above the crossover Mg concentration.

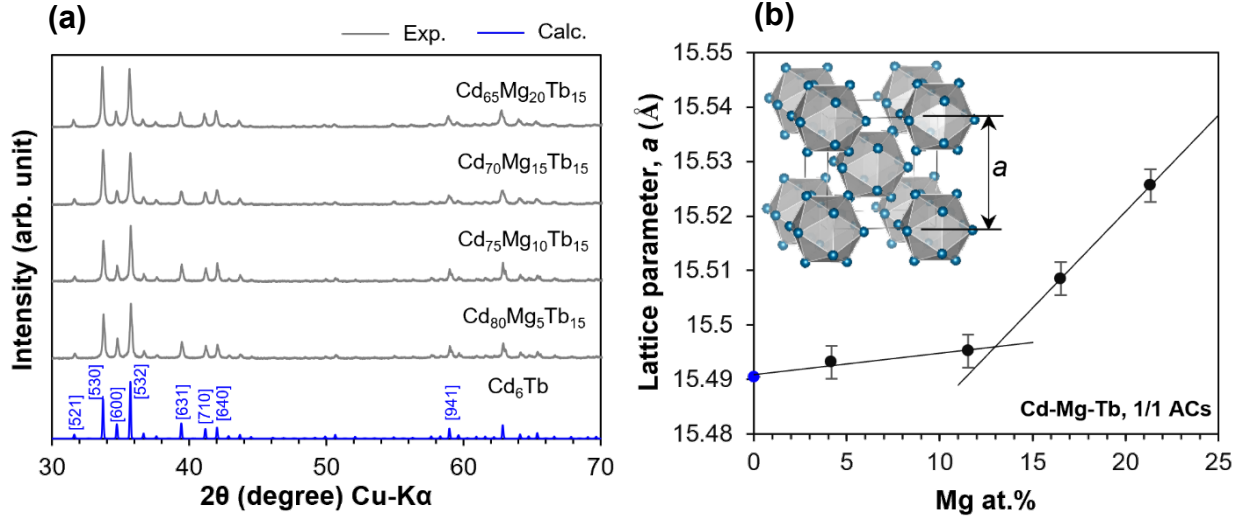

**Figure 3.** (a) Powder x-ray diffraction (XRD) patterns of the Cd-Mg-Tb 1/1 APs containing 5, 10, 15 and 20 at. % Mg after annealing at 773 K. The calculated XRD pattern obtained from the refined structure of the binary $Cd_6Tb$ [23], a blue pattern, is co-plotted on the bottom of the same figure for comparison. (b) Variation of lattice parameters of the cubic 1/1 APs as a function of their constituent Mg content. Here we used the analyzed Mg composition.

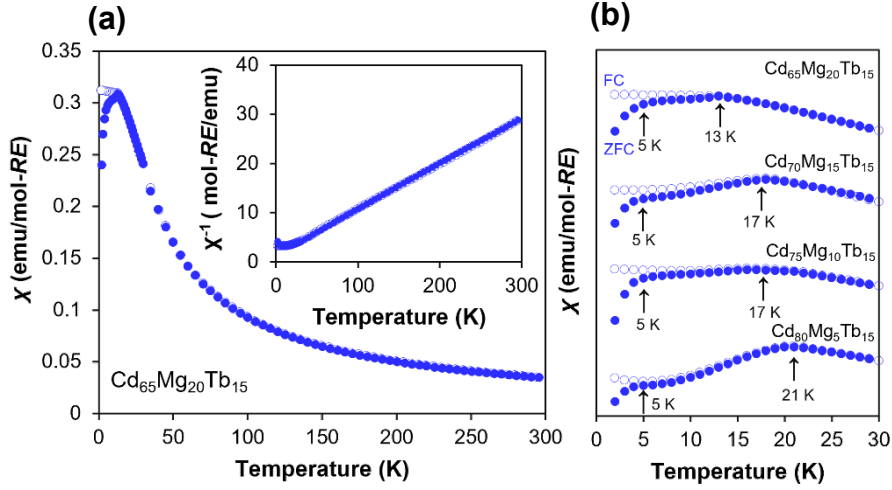

**Figure 4.** (a) Temperature dependence of DC magnetic susceptibility of the Cd$_{65}$Mg$_{20}$Tb$_{15}$ 1/1 AP measured under $H$ = 100 Oe in the temperature range of 2–300 K showing both field cooled (FC) and zero field cooled (ZFC) behaviours. Open and close circles represent FC and ZFC curves, respectively. The inset shows the inverse susceptibility of the same alloy. (b) Magnified low-temperature DC magnetic susceptibilities of the 1/1 APs containing 5, 10, 15 and 20 at.% Mg in Cd-Mg-Tb system.

**Table 2.** Weiss temperature ($\Theta_w$), the freezing temperatures ($T_{f1}$, $T_{f2}$), and effective magnetic moment ($\mu_{eff}$) of the Cd$_{85-x}$Mg$_x$Tb$_{15}$ ($x$ = 5 and 20) 1/1 APs. Appearance of successive phase transitions is associated with the existence of various antiferromagnetic magnetic configurations, which are nearly degenerate, over a bcc array of icosahedral clusters.

| Composition | $\mu_{eff}$ ($\mu_B/RE_{ion}$) | $\mu_{calc.}$ ($\mu_B/RE_{ion}$) | $\Theta_w$ (K) | $T_{f1}$ (K) | $T_{f2}$ (K) | $T_N$ (K) |
|---|---|---|---|---|---|---|
| Cd$_{80}$Mg$_5$Tb$_{15}$ | 9.6±0.2 | 9.72 | -13.7±1.5 | 5.0±0.5 | – | 20.5±0.5 |
| Cd$_{65}$Mg$_{20}$Tb$_{15}$ | 9.4±0.2 |  | -17.6±1.5 | 5.0±0.5 | 13.5±0.5 | – |

In order to gain further insights into the AFM to SG-like magnetic transition from the viewpoint of the crystal structure, TEM experiments have been carried out. Figure 6 presents the selected area electron diffraction (SAED) patterns of the Cd$_{80}$Mg$_5$Tb$_{15}$ and Cd$_{65}$Mg$_{20}$Tb$_{15}$ 1/1 APs taken with an incidence along [111]$_C$ (subscript C denotes a cubic lattice) at $T$ = 100 K and 300 K. While the SAED pattern of Cd$_{80}$Mg$_5$Tb$_{15}$ at $T$ = 100 K exhibits superlattice reflections at the midpoints of the fundamental reflections, as marked with red arrowheads in Figure 6(a), that of the same sample at $T$ = 300 K shows no superlattice reflections (see Figure 6(b)). The superlattice reflections in Figure 6(a) indicate the doubling of the periodicity along [101]$_C$, clearly breaking the threefold symmetry. This means that a structure phase transition has been occurred in the Cd$_{80}$Mg$_5$Tb$_{15}$ during cooling. In fact, similar phenomena have been commonly observed in the binary Cd$_6$RE and Zn$_6$Sc 1/1 APs, and have been described as order-disorder-type phase transitions associated with the orientations of the Cd$_4$ tetrahedra. Here, the low temperature phase takes a monoclinic structure [11,15,27,29–32]. In the binary systems, the transition temperature depends on the RE type, but is always in the range of $100 < T < 200$ K [29,30,33]. Figure 7 presents a schematic illustration of the low-temperature superstructure of the Cd$_{80}$Mg$_5$Tb$_{15}$ 1/1 APs with respect to orientational ordering of the Cd$_4$ tetrahedra. In the figure, planes with different colors represent different orientations of the Cd$_4$ tetrahedron. Solid lines indicate cubic unit cells for the high-temperature phase, while dashed lines show a monoclinic unit cell for the low-temperature phase. Interestingly, the SAED patterns of the Cd$_{65}$Mg$_{20}$Tb$_{15}$ 1/1 AP at both $T$ = 100 and 300 K (Figures 6(c) and 6(d)) show no superlattice reflections down to low temperatures indicating that, unlike Cd$_{80}$Mg$_5$Tb$_{15}$, the crystal remains disordered in terms of the orientations of the Cd$_4$ tetrahedra. The absence of the ordering in the orientations of the tetrahedra might possibly explain why Cd$_{65}$Mg$_{20}$Tb$_{15}$ exhibits no AFM ordering but only SG-like transition at low temperatures (see Figures 4 and 5). To the best of our knowledge, this is the first observation that indicates high Mg content of the 1/1 AP hinders the ordering of central $Cd_4$ tetrahedra and further long-ranged AFM ordering.

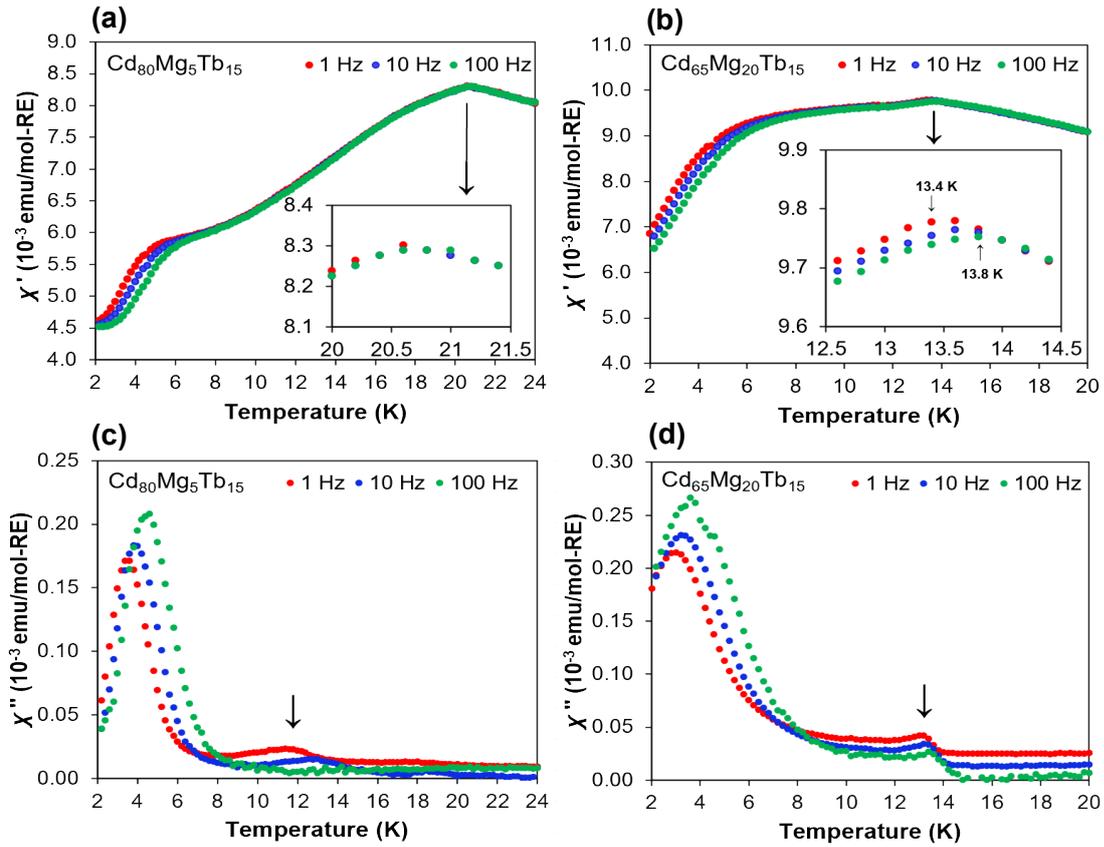

**Figure 5.** The in-phase and out-of-phase components of the AC susceptibilities for the (a,c) $Cd_{80}Mg_5Tb_{15}$ and (b,d) $Cd_{65}Mg_{20}Tb_{15}$ measured under $f_{AC}$ = 1 - 100 Hz. The insets in Figures 5(a) and 5(b) show the magnified view of the susceptibility curves around the cusp.

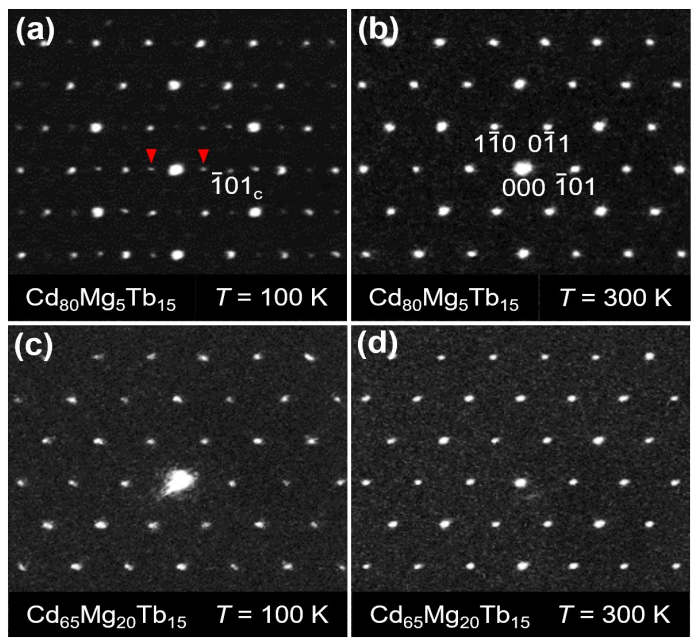

**Figure 6.** Selected area electron diffraction (SAED) patterns taken with an incidence along $[111]_C$ of the (a,b) $Cd_{80}Mg_5Tb_{15}$ and (c,d) $Cd_{65}Mg_{20}Tb_{15}$ at 100 K and 300 K. The superlattice reflections are found in the pattern of the $Cd_{80}Mg_5Tb_{15}$ at 100 K, as indicated by arrow heads.

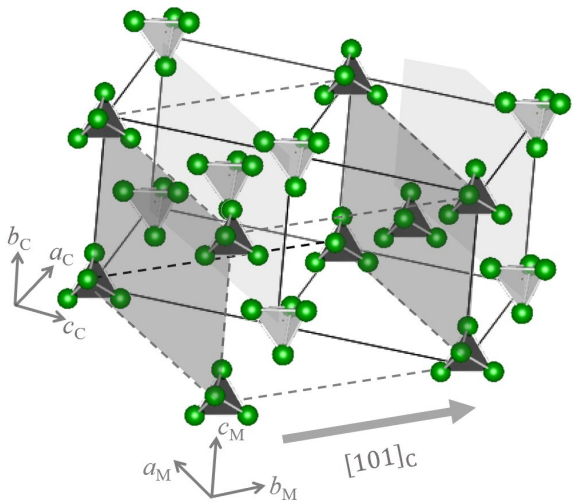

**Figure 7.** Schematic illustration of the possible low-temperature superstructure of $Cd_{80}Mg_5Tb_{15}$ regarding the orientation of the $Cd_4$ tetrahedron at the center of the each RTH cluster. Planes with different colours represent different orientations of the $Cd_4$ tetrahedron. ($a_C$, $b_C$, $c_C$) and ($a_M$, $b_M$, $c_M$) stand for a cubic and monoclinic lattice basic vectors, respectively.

One possible factor that hinders the orientational ordering of $Cd_4$ tetrahedron in alloys with high Mg content may be the effect of Mg atoms that enters into the central tetrahedral shell. Based on the only structure analysis in hand, i.e. the structure of the $Cd_{65}Mg_{20}Y_{15}$ 1/1 AP shown in Figure 8 [34], the three most preferred Mg occupational sites are the vertex positions of the rhombic triacontahedron (M6 sites), the positions on the two-fold axes of the icosidodecahedron (M5 sites) and the central tetrahedron (M7 sites), with 87%, 72% and 72% occupational fractions of Mg, respectively. The M6 sites, in particular, shows a relatively strong tendency to accept Mg. The numbers in parentheses in Figure 8 refer to Mg occupational fractions. It is instructive to note that the Mg content that resides on the M6 sites alone amounts to 12.42 at.% in the total composition. The latter value is reasonably close to the crossover point of Mg concentration in Figure 3(b) (i.e. ~ 13 at.%). This suggests a possible scenario that upon increasing the Mg content, most Mg atoms would first go into the M6 sites if the Mg content is low, while if Mg content exceeds ~ 13 at.%, Mg atoms would start to occupy the other sites including those in the central tetrahedra, inhibiting the ordering of their orientations. This picture needs to be further verified through more detailed analyses, which is left for future studies.

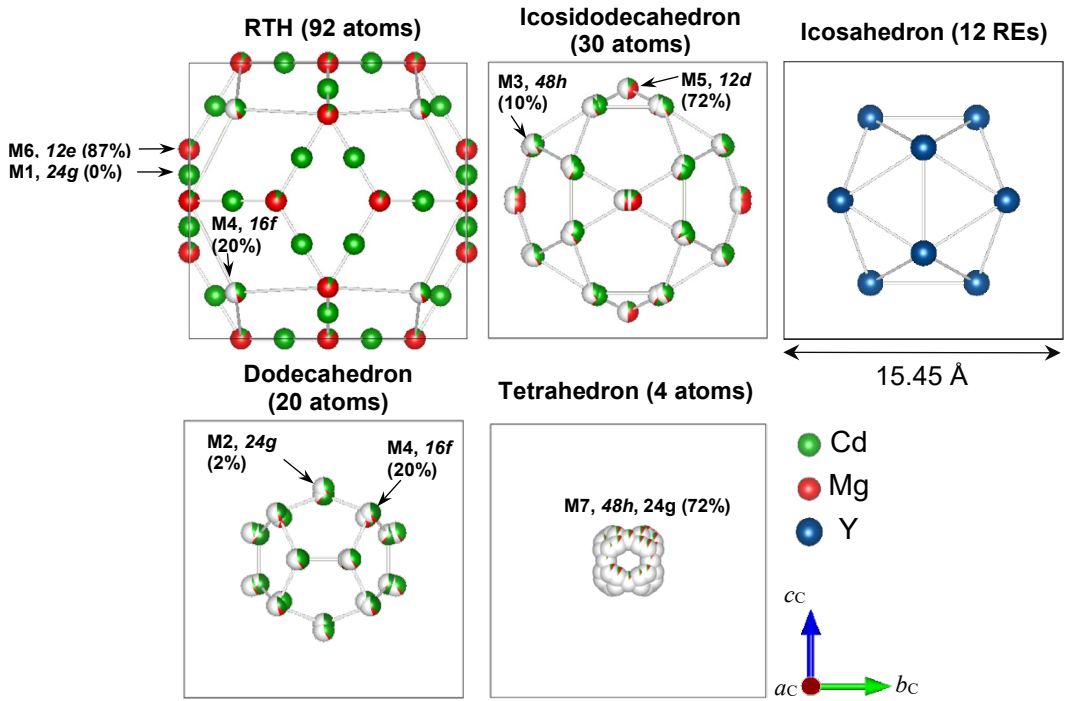

**Figure 8.** The successive sequence of atomic shells in the $Cd_{65}Mg_{20}Y_{15}$ 1/1 AP [33]. Occurrence of 60 atoms in the central tetrahedron is due to the two types of disorders: a 90° rotational disorder along its 2-fold axis and a triple splitting of the tetrahedral corner positions. The lattice parameter is 15.45 Å. The numbers in parentheses refer to Mg occupational fraction.

Another important factor to consider when discussing the order-disorder transition in 1/1 APs in the course of increasing Mg content is the size effect of the second shell dodecahedra that encage the central tetrahedra. It has been commonly assumed that too short interatomic distances between the central tetrahedron and the surrounding dodecahedron would induce a deformation of the dodecahedral shell thus costing elastic strain energy, while the ordering would take place to minimize the energy [29,35]. At high-enough temperatures, the entropy gain could compensate for the energy cost. Using the Helmholtz formula, the differential free energy $\Delta F$ upon disordering can be written as $\Delta F = \Delta E - T \Delta S$, where $\Delta E$ (>0) and $\Delta S$ (>0) denote the energy and entropy differences, respectively, (i.e. $\Delta E = E_{disordered} - E_{ordered}$, $\Delta S = S_{disordered} - S_{ordered}$) at temperature $T$. As a crude approximation, one can assume that $\Delta S$ is a constant. Then the transition temperature $T_c$, which follows from $\Delta F = \Delta E - T_c \Delta S = 0$, scales linearly with $\Delta E$. Importantly, as the lattice parameter increases, the volume of the dodecahedron expands and the distances between the neighboring atoms located on the tetrahedron and the dodecahedron increases, which leads to lower strain energy and thus lower $\Delta E$. For Cd$_6$RE 1/1 APs having larger RE atoms and thus a larger volume of the dodecahedron, the transition temperature is reported to be relatively lower than those with smaller RE size [29,35]. The present argument is fully consistent with this observation, and can also be naturally extended to the present system of ternary alloys. Note that the dodecahedron volume increases following the significant rise in the lattice parameter of 1/1 APs above ~ 13 at.% Mg (see Figure 3(b)) leading the structural transition temperature ($T_c$) to drop. In this scenario, the structural transition in Cd$_{65}$Mg$_{20}$Tb$_{15}$ 1/1 AP (nominal composition) may still take place below $T = 100$ K. However, the probability of the tetrahedron reorientation is proportional to $\exp(-E_a/k_B T)$, where the $E_a$ and $k_B$ are the activation energy of the tetrahedron reorientation and the Boltzmann factor, respectively [29,35]. In this sense, the reorientation of the tetrahedron (disordering) is strongly dependent on the temperature and might be kinetically frozen at lower temperatures.

Here, the underlying reasons of the composition-driven AFM to SG-like magnetic transition will be discussed. The composition-dependent AFM to FM and FM to SG transitions have been reported in the Au-Al-RE (RE = Gd and Tb) systems [26,36] and been interpreted by the impact of Au/Al variation on the electron-per-atom (*e/a*) ratio of the compounds. However, given that both Mg and Cd are divalent, the *e/a* values of the present Cd$_{85-x}$Mg$_x$Tb$_{15}$ (*x* = 5, 10, 15, 20) compounds equal 2.15. Therefore, the AFM to SG-like magnetic transition in the present study is perhaps not triggered by *e/a* ratio, but rather by the suppression of Cd$_4$ tetrahedra ordering, which is a key element in the establishment of AFM magnetic transition. Indeed, the cubic to monoclinic structural transition induces a strong distortion which plays an important role in relieving some degree of magnetic frustration inherent to the RE elements with nearly perfect icosahedral arrangement in the cubic structure and favors the occurrence of AFM magnetic transition [27]. Nevertheless, the structural disorder of inner Cd$_4$ tetrahedron may induce randomness of crystal electric field (CEF) and raise the SG-like freezing of the RE magnetic moments at low temperatures. The present result calls for further experimental and theoretical studies for more clarification.

Note that the chemical disorder of the Cd/Mg should also be taken into account as the source of disorder contributing to the observed spin-glass behavior in alloys with high Mg content. The importance of chemical disorder was recently confirmed by measuring CEF splitting in Au-Si-Tb 1/1 AP [37]. It seems reasonable, therefore, to assume that the combination of chemical disorder due to a randomized substitution of Cd with Mg and the orientational disorder of the Cd$_4$ tetrahedra is responsible for the occurrence of the AFM to SG-like magnetic transition in the present compounds.

## 4. Conclusion

This research was undertaken to evaluate the effect of Mg/Cd substitution on the crystal structure and magnetic properties in the Cd$_{85-x}$Mg$_x$Tb$_{15}$ (*x* = 5, 10, 15, 20) 1/1 APs. Each of the four compounds exhibited two anomalies in their DC magnetic susceptibilities. The higher transition temperature was found to be of the AFM-type in the Cd$_{80}$Mg$_5$Tb$_{15}$ and SG-type in the Cd$_{65}$Mg$_{20}$Tb$_{15}$. The superlattice reflections are observed in the SAED pattern of the Cd$_{80}$Mg$_5$Tb$_{15}$ at $T = 100$ K, whereas no superlattice reflections are noticed for the Cd$_{65}$Mg$_{20}$Tb$_{15}$ at the same temperature. The appearance of superlattice reflections is associated with the order-disorder-type phase transition with respect to the orientations of the Cd$_4$ tetrahedra. Taken together, the combined effect of chemical disorder of Cd/Mg and orientational disorder of the Cd$_4$ tetrahedra is presumably responsible for the occurrence of the AFM to SG-like magnetic transition in the present compounds.

### Acknowledgements

This work was supported in part by Japan Society for the Promotion of Science through Grants-in-Aid for Scientific Research (Grant Nos. JP19H05819, JP17K18744, JP19H01834, JP19K21839, JP19H05824, JP19K03709 and JP18K13987). This work was also supported partly by the research program "dynamic alliance for open innovation bridging human, environment, and materials".